# Assessing the properties of a colloidal suspension with the aid of deep learning

Tomasz Jakubczyk [a], Daniel Jakubczyk [b,*], Andrzej Stachurski [a]

[a] Institute of Control and Computation Engineering, Warsaw University of Technology
ul. Nowowiejska 15/19, PL-00665 Warsaw, Poland
[b] Institute of Physics, Polish Academy of Sciences
al. Lotników 32/46, PL-02668 Warsaw, Poland

**Abstract**

Convolution neural networks were applied to classify speckle images generated from nanoparticle suspensions and thus to recognise suspensions. The speckle images in the form of movies were obtained from suspensions placed in a thin cuvette. The classifier was trained, validated and tested on both single component monodispersive suspensions, as well as on two-component suspensions. It was able to properly recognise all the 73 classes – different suspensions from the training set, which is far beyond the capabilities of the human experimenter, and shows the capability of learning many more. The classes comprised different nanoparticle material and size, as well as different concentrations of the suspended phase. We also examined the capability of the system to generalise, by testing a system trained on single-component suspensions with two-component suspensions. The capability to generalise was found promising but significantly limited. A classification system using neural network was also compared with the one using support vector machine (SVM). SVM was found much more resource-consuming and thus could not be tested on full-size speckle images. Using image fragments very significantly deteriorates results for both SVM and neural networks. We showed that nanoparticle (colloidal) suspensions comprising even a large multi-parameter set of classes can be quickly identified using speckle images classified with convolution neural network.

1. **Introduction**

A need for quick assessing of the (selected) properties of colloidal suspensions, in particular in microdroplets, has recently become quite obvious in the context of biologically active anthropogenic aerosols, in particular carrying viruses. It is widely accepted that the analysis of coherent light scattering – both spatial and temporal – can provide a versatile non-contact tool for such assessment (see e.g. [1–7] and references therein). When a coherent wave is scattered by a randomly inhomogeneous medium, like a suspension, a so called speckle pattern forms. The pattern is three-dimensional and as long as the distribution of scatterers in the scattering medium is stationary, so is the speckle pattern. However, if the scattering medium evolves somehow, the scattering pattern is dynamic and in a quite complicated manner. Being an interferogram, it is extremely sensitive to the distribution of scatterers and the parameters of illuminating wave (wavelength, polarisation, wavefront curvature). In consequence, the pattern carries significant information on the scattering system. Again, similarly as for other interferograms, any of its fragments, e.g. a part of a section, as an image obtained on a sensor, also carries much of this information. The smaller the fragment, the less detailed information.

It is however not obvious how to extract the encoded information. There is no known procedure to solve such an inverse problem, though there exist statistical tools for handling speckles [8,9], which can help in some cases. An alternative approach seems however possible – the scattering images could be directly classified with machine learning (ML)[10,11]. Since the algorithms in classification/recognition tools provided with ML are hardly comprehensible for humans, efforts are made to interpret/visualize their operations (see e.g.: [12,13] and references therein).

[*] Corresponding author. E-mail address: Daniel.Jakubczyk@ifpan.edu.pl

In this work, instead of solving an inverse problem, we've tried to harness convolution neural networks (CNNs) for assessing the colloidal suspension properties by classifying characteristic light intensity patterns (speckles) generated due to scattering.

Speckles often constitute an unwanted noise plaguing an image. Then, appropriate exploiting their properties enables restoring a clear image [14–17]. On the other hand, there are also promising applications of speckle patterns created on purpose [18–20].

Since a human observer/experimenter cannot easily grasp the relations between the scattering system properties and the speckle pattern, hence the idea of harnessing artificial neural networks. If properly designed and trained, they can often process the information given (e.g. noisy images [14]) or abstract/recognise useful information from a human-unintelligible set essentially without constructing an explicit model. It seems that so far the number of abstracted classes has been usually limited to several [21–24].

In this work, we want to test the capabilities of the neural networks for recognizing more numerous classes of scattering patterns and possibly generalising its predictions.

If the method turns out successful, in the next step we would apply it to scattering on microdroplets of suspensions, which, so far, we have studied with other non-contact (mostly optical) methods (see e.g. [5,25,26]). The far-reaching objective is to devise a quick diagnostic tool for suspension microdroplet characterisation, which eventually would allow identifying microorganisms/viruses present in microdroplets. In the days of COVID-19 pandemic, such application appears very welcomed.

## 2. Experiment

In this experiment, we tried to constrain the number of classes that the neural network would have to recognize, and thus we used a very simple scattering system with a limited number of degrees of freedom – see Fig. 1. It is meant to introduce as little experimental conditions variation from run to run as possible.

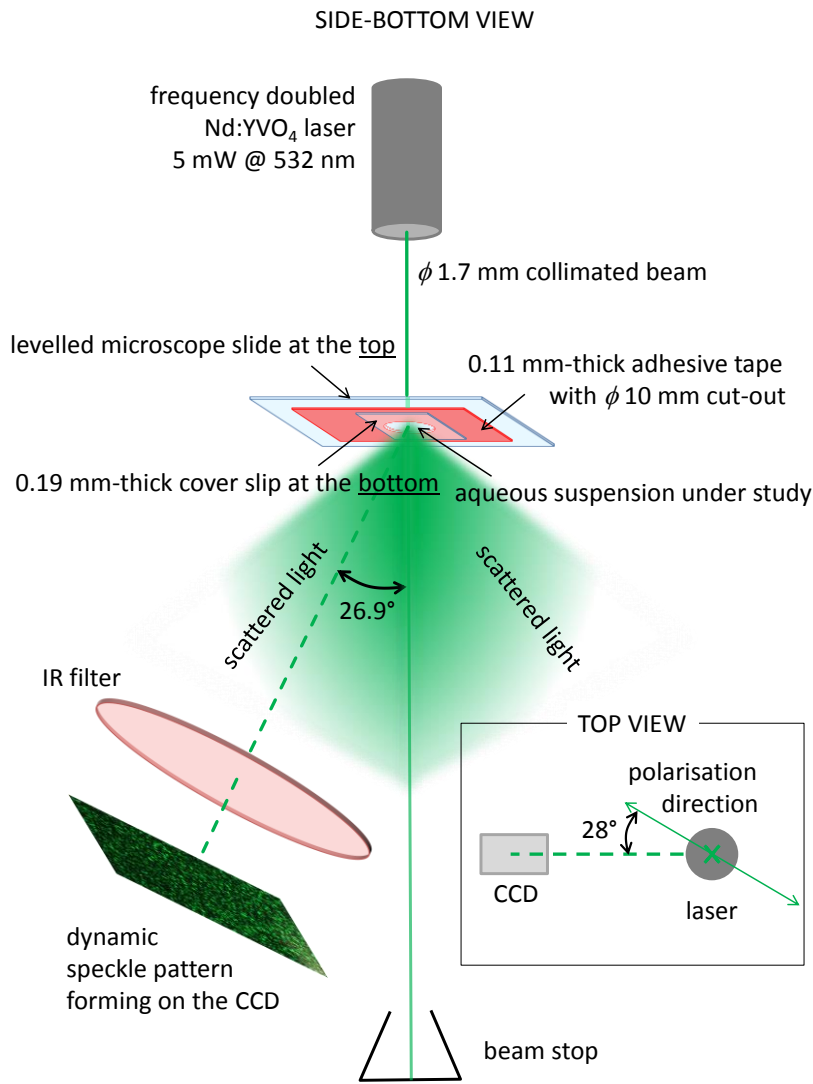

Figure 1. Experimental setup schematic drawing.

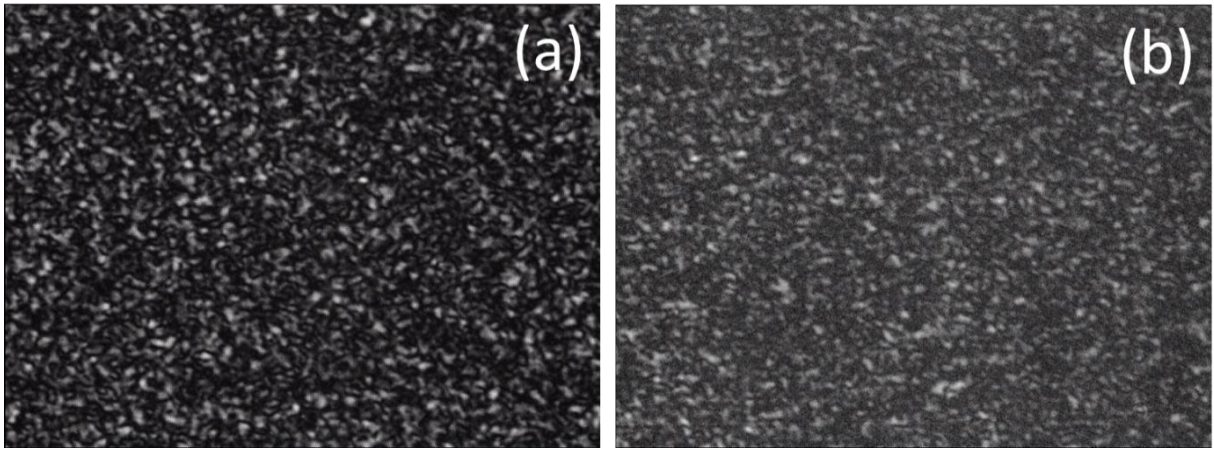

Figure 2. Sample movie frames corresponding to two very different substances: (a) polystyrene microspheres (PS) 1100 nm diameter, (b) ultrapure water. The images are brightness-corrected only – no background subtracted. In such form, the static images are hardly distinguishable (apart from the brightness of the background). The movies, however, appear quite different. In *PS 1100 nm* movie, the bright speckles are clearly in motion, while for ultrapure water, the movie seems static, apart for a very faint noise in the background.

An aqueous (colloidal) suspension was loaded into a flat, 0.11 mm thick cuvette, which was illuminated perpendicularly with a laser beam. The cuvette thickness was chosen to minimise multiple scattering, but also to enable observation of speckles from weakly scattering suspensions. The cuvette was carefully levelled in order to suppress the movement of the suspended particles due to macroscopic flow. The experiment was conducted at room temperature, which was stabilised at 21.5±1°C with air-conditioning system. The observation angle was chosen to match the velocity of speckle movement with the capabilities of the camera, but to avoid strong forward scattering from stationary scatterers within the light beam (on/in optical glasses). Much care was taken to avoid static scattering due to cuvette contamination. The illuminating beam polarisation was constant but randomly chosen. Since a green Nd:YVO$_4$ laser utilizes a 2-step conversion with limited efficiency from infrared, an IR filter had to be placed before the CCD detector to observe only speckles at 532 nm. On top of that, only the green channel from the camera was used. We used a 14-bit colour camera (Pike F-032C, AVT) and wrote a dedicated codec to make full use of its high dynamic range. The coherence condition (laser-sample-detector distance) was fulfilled for maximum speckle contrast.

For each scattering sample, a 500-frame movie was recorded at 80 fps (see sample movies in [27] and sample frames in Fig. 2). Since scattered light intensity differed by several orders of magnitude between the samples, the exposition (and if necessary – the camera gain) had to be adjusted. Only several discrete values of these were used. In case of doubt, whether the dynamic range was fully utilised, the measurement was repeated for a neighbouring exposition/gain value. Changing the camera parameters between the experimental runs – in particular, the exposition (image integration time) and gain (also amplifying the electronic noise) – may introduce non-meritorical information into the pictures and mislead/hinder the learning. Some difference of brightness between movies (series of images) seems unavoidable and was coped with at the data preparation stage.

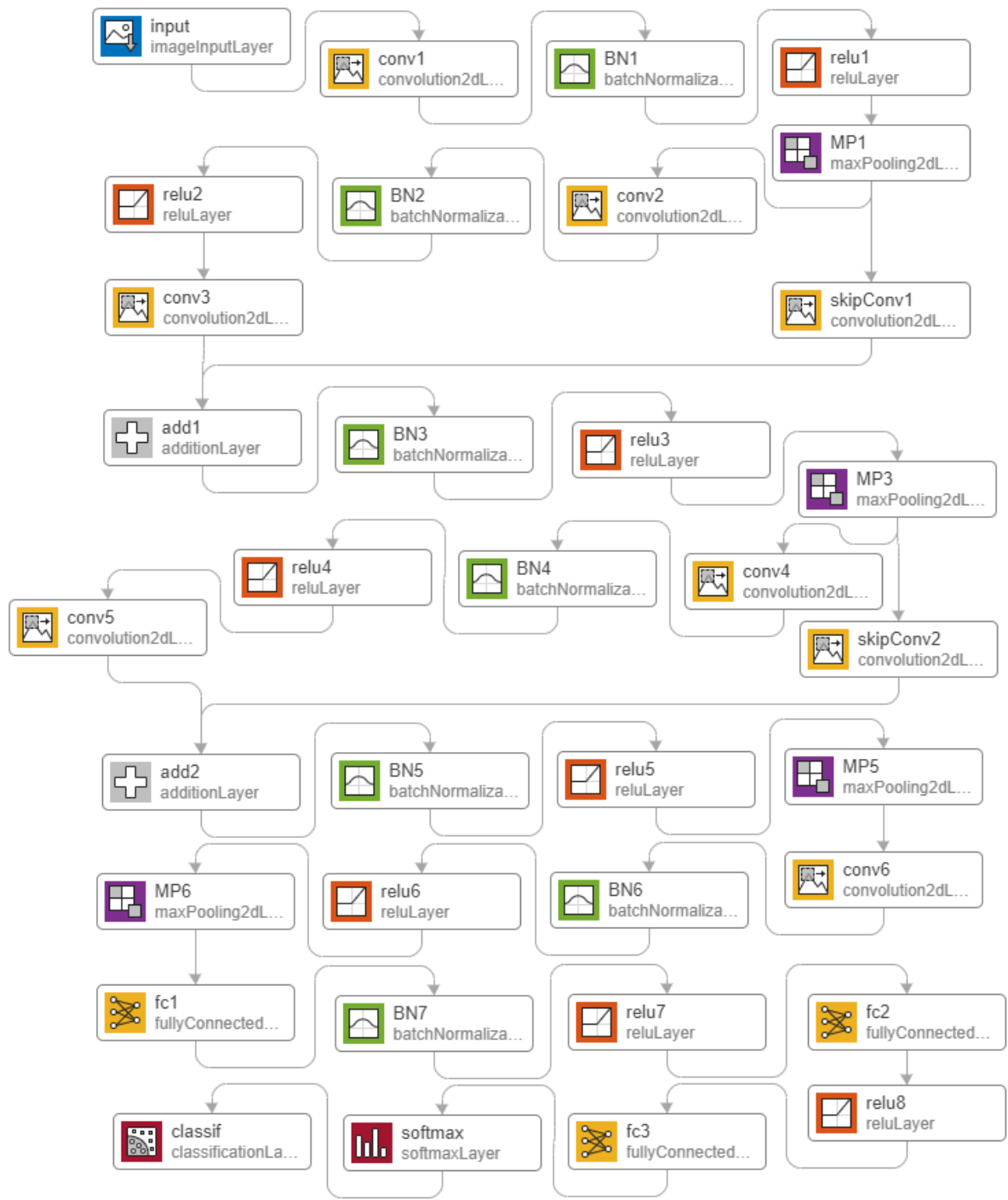

Figure 3. Graphic representation of the structure of the final neural network.

For network training/validation/testing we first used a set of aqueous suspensions of spherical (nano)particles covering a wide range of diameters from 10 nm to 3 $\mu$m (20 sizes). There were 5 substances among the nanoparticle (NP) material: Au, Ag, $SiO_2$, $TiO_2$ and polystyrene (PS), which corresponds to 5 different refractive indices. Refractive index of NPs is usually somewhat different than that of material in bulk and often depends on the particle size (and shape) – in particular for metallic NPs (see e.g.[28]). In the presented experiment, the exact value of the refractive index was not important and did not have to be established. For reference we provide indicative values corresponding to our experimental conditions (532 nm, 21.5°C): Au and Ag in 300-nm-thick film [29] – 0.424 + 2.33i and 0.0425+3.45i respectively, fused silica

NPs [30] – 1.461, TiO$_2$ 35-nm NPs (anatase phase) in water [31] – 2.54 and PS microspheres in water [32] – 1.588. Several types of suspension were provided in 8 dilutions (roughly logarithmic series) of the base suspension. Since the base concentration of NPs in different suspensions was sometimes much different (from 5.46×10$^{-3}$ to 15 wt%) the absolute NPs concentration can hardly be and was not compared in this experiment. The dilutions were prepared with distilled water. We also used ultrapure water (Simplicity, Millipore/Merck) and tap water as references. The testing was later performed with several binary mixtures of the above suspensions.

We have recorded 639 files (183 GB total), which corresponds to ~ 3.2×10$^5$ scattering images. The images for training, validation and testing were randomly selected from each class in 8:1:1 proportion respectively.

For the sake of machine learning, individual frames of movies – files – were assigned (surjection) to specific classes (obviously, all the frames of a single movie belong to the same class). The name of a class consisted of the suspended phase substance name/symbol, concatenated with the suspended particles size and the dilution factor of the suspension.

## 3. Neural network architecture

Several neural network architectures were tested. The architectures were devised in an iterative manner, using a platform provided by MATLAB (version 020a) – all the tested architectures can be found in [33]. The key tool used was Deep Learning Toolbox with CUDA framework capability. The calculations were performed on a PC-class computer equipped with *GTX TITAN Black* GPU (nVidia). The results of training the network – successes and failures – were analysed and the conclusions were implemented in the following version of the network. We also performed experiments and comparisons with the Support Vector Machine (SVM) – see Appendix A. We also intended to analyse a sequence of scattering images (movie) at once. Speckle dynamics carry information on the suspended particles hydrodynamic radius. Dynamic light scattering technique (DLS) [34,35] extracts this information by analysing temporal correlations of the scattering signal. Though our CCD camera was not fast enough for a proper DLS, we expected that we could further improve classification and abstracting capabilities of our networks. The project however turned out to significantly exceed our hardware – we estimate that at least 20 GB of memory would be required – and had to be postponed.

The architecture that we found most satisfactory so far (labelled as *t7g24* in [33]) is presented in Fig. 3. The optimisation algorithm was set to Stochastic Gradient Descent with Momentum (SGDM), where momentum = 0.9 (default). Several hyperparameters were tuned manually in the process of network designing and training (see Tab. 1), while the rest was left default as described in the Deep Learning Toolbox documentation [36].

| Hyperparameter | Value |
| --- | --- |
| *InitialLearnRate* | 0.01 |
| *Shuffle* | every-epoch |
| *ValidationFrequency* (its value is a trade-off between long validation time and frequency of validation) | 1000<br>4000 in 11$^{th}$ Epoch |
| *ValidationPatience* (changed to stop training when the network stops to learn further) | 5 |
| *LearnRateSchedule* | piecewise |
| *LearnRateDropPeriod* (often modified so that the learning step changes when it is beneficial for training) | 10 |
| *MiniBatchSize* (selected as large as possible, but not to exceed the memory available for the network – 5 GB) | 16 |

Table 1. The hyperparameters of the network training, which were manually tuned in the experiment.

The input layer was 640×480 pixels, which corresponds to the size of the movie frame. As mentioned above, only one colour channel was selected – green – corresponding to the colour of the laser light.

There are two factors, which easily become predominant features determining the membership of a class: the image/movie brightness – mentioned in the "Experiment" section – and stationary background of the speckle (interference) pattern, originating from the light scattering on the stationary system elements (in particular, on the stationary impurities). Such false classification must be avoided as non-meritorical. To overcome the difficulty and prepare good training data, we processed the obtained raw images in the following way.

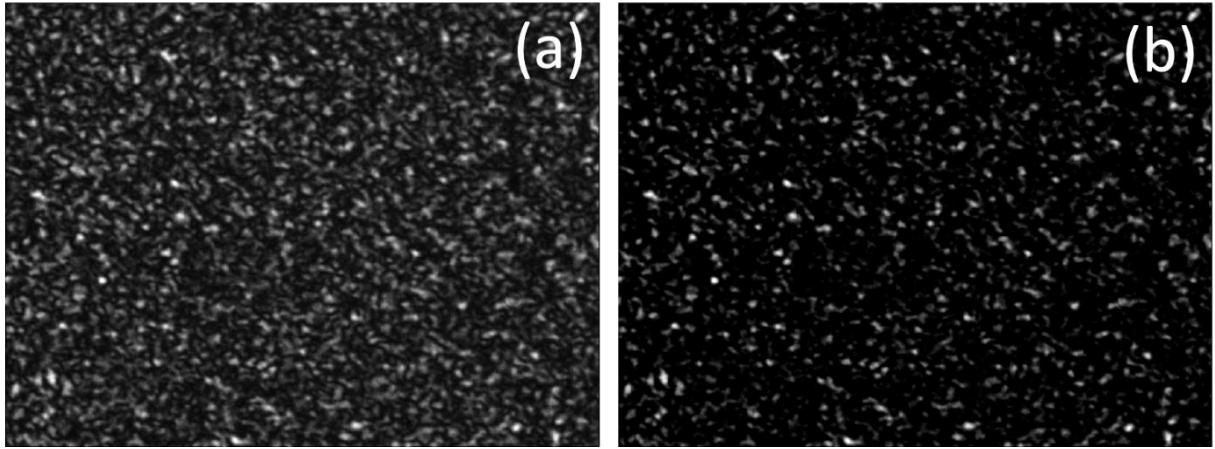

Figure 4. A sample movie frame corresponding to PS 1100 nm diameter: (a) raw image (b) image with the background subtracted. The speckles corresponding to scattering on static elements are removed in this way. Both images are brightness-corrected in the same degree for clarity.

An average frame was calculated for each consecutive 100 frames of the movie. Such average carries enhanced information about scattering on stationary system elements, as the scattering on moving particles gets randomised. The obtained average was then subtracted from each frame to suppress the effect of stationary scatterers and enhance speckles from moving NPs – see Fig. 4. The operation also generally raises the contrast of the speckle pattern.

The brightness of each obtained pattern was normalised to 0-1 range. This operation was found highly beneficial. It suppressed the influence of the (raw) image brightness upon its classification and, even more importantly, it significantly augmented numerical stability of the learning algorithm.

The network consists of 8 convolution layers: *conv1-6* and *skipConv1-2*. Each consists of 64 3×3 sized filters. The final filter size has been selected as a result of trials with earlier versions of the network. It is generally accepted that several layers of smaller filters work the same or better than a single layer of larger filters. In order to match the size of the output images between the layers, different values of *stride* and *padding* parameters were selected. *Stride* in layers *conv1*, *conv3*, *conv4* and *conv6* has been set to default value of 1 in both directions. Whereas in the *conv2*, *conv5*, *skipConv1*, *skipConv2* layers the *stride* parameter has been set to 2 in both directions, which has the effect of reducing the output image size, similar to *Pooling* layers. The size 1 padding, in both directions, was added in the *conv2*, *conv3*, *conv4* and *conv5* layers to prevent the image from resizing due to the application of filters. The *skipConv1* and *2* layers are arranged parallel to *conv2-3* and *conv4-5* layers respectively.

There are 4 *maxPooling* layers (*MP1*, *MP3*, *MP5*, *MP6*) with 3×3 filter size, moved with *stride* 2 in both directions. This reduces the size of the input image by half in both directions, while preserving the strongest activation within the filter for each point of its application. 7 *batchNormalization* layers helping to maintain good numerical stability of the learning algorithm are located directly behind the *conv1*, *conv2*, *add1*, *conv4*, *add2*, *conv6* and *fc1* layers. Their placement results from the tests carried out on previous versions of the network, with the main criterion being to ensure effective network learning.

The 2 fully connected layers (*fc1*, *fc2*) are among the last layers of the network and are designed to perform the necessary non-linear transformation of the features extracted by the convolution layers into a vector that allows the network to determine the affiliation of a given image to a set of classes. The *fc1* and *fc2* layers consist of 512 and 256 neurons respectively. This choice of the number of neurons in fully interconnected layers aimed to provide a sufficient number of parameters to allow the network to classify effectively, while maintaining good generalization capabilities, a reasonable time of network learning and obtaining a classification result, but above all to fit into the operating memory of the graphics card.

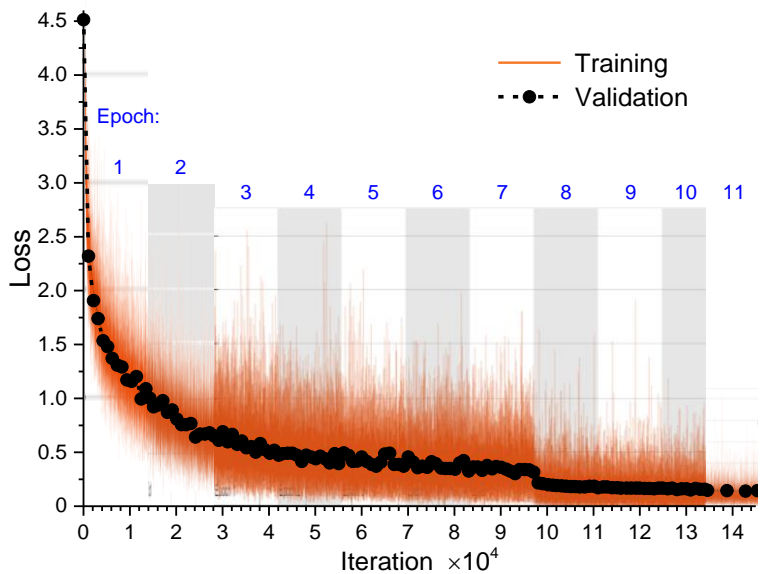

Figure 5. The loss function evolution during training of *t7g24* network on single- and two-component suspensions at once. The training lasted 162 hours and was performed in 4 sequences: epoch 1, 2, 3-10 and 11. After the epoch 7, the *LearnRate* was decreased 10-fold. The 3$^{rd}$ sequence was interrupted prematurely because of power failure. In order to verify whether the training could be considered concluded, it was restarted from a check-point as the epoch 11 with the *ValidationFrequency* parameter 4-fold lower.

As the depth of the network increases, the so-called vanishing gradient problem appears [37]. It manifests as a decrease or complete disappearance of the network's learning progress. This is due to repeated – for each layer – numerical calculation of derivatives (gradient), so that small rounding related to the accuracy of the floating point representation of numbers eventually leads to the loss of information necessary for further learning of subsequent layers of the network.

From previous experiments, it was concluded that the input image size must be reduced with the use of convolution layers (some layers can reduce the image size with stride > 1) and *max Pooling* to a level where the number of weights in the first fully connected layer will be small enough to conveniently fit into the global graphics card memory. However, it turned out that an attempt to reduce the size excessively (> 2×) in a single convolution layer leads to a significant loss of information, which results in the cessation of network learning. This means that it is necessary to build a convolution network with the appropriate depth – about 6 layers reducing the image size by half each.

The *add* and *skip* layers are components of the architecture that work on a similar basis to *ResNet* [38]. In our case, the skip layers are fed from *MP1* and *MP3* layers and are designed to provide the same image size reduction as the adjacent layers. The add layers sum up the input images into one of the same size. We've established that adding *skip* and *add* layers had a positive effect on network learning in the presented case. We also experimented with *dropout* layers, as potentially beneficial, but finally didn't use them. *Dropout* consists in random zeroing of inputs according to a specified probability.

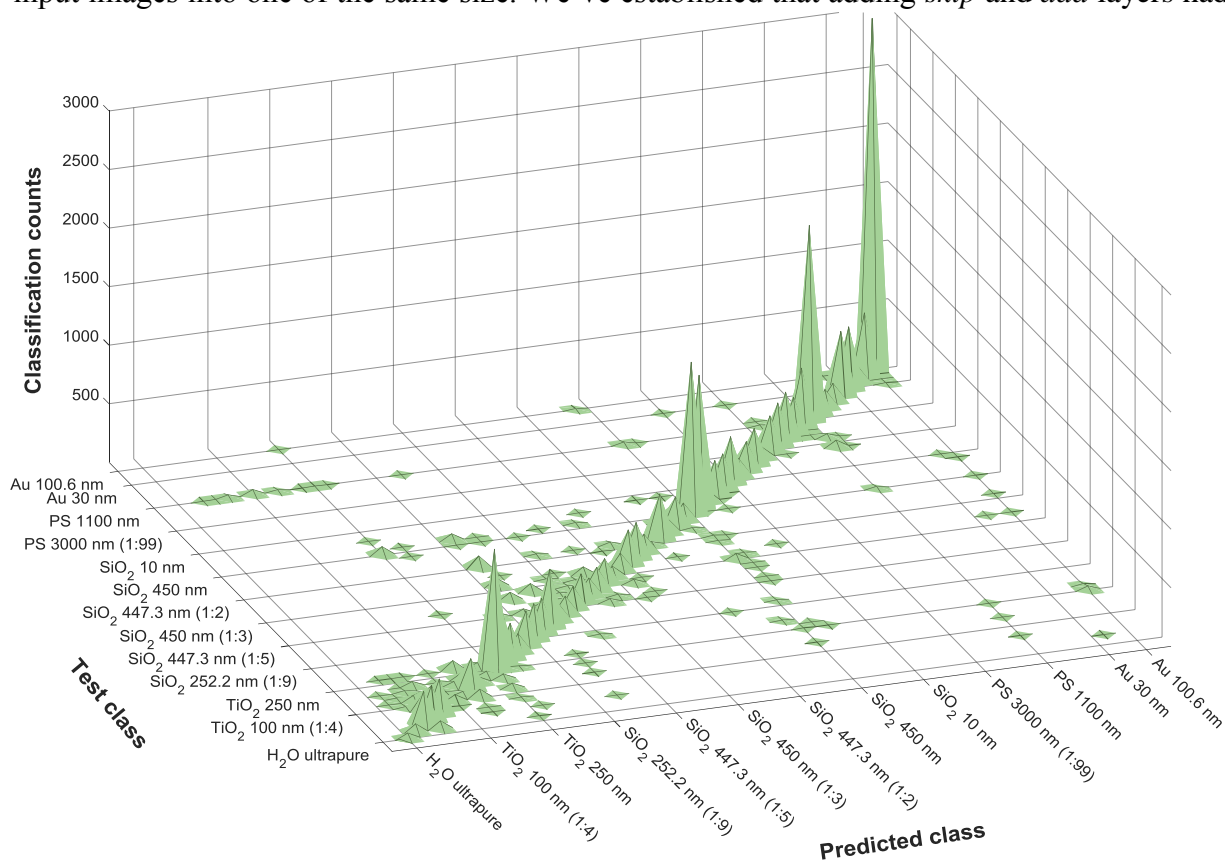

Figure 6. Graphic representation of the confusion matrix for the final network trained on single-component colloidal suspensions, ultrapure and tap water. Only every 5-th class label shown for clarity. Each label states the substance of the suspended phase, suspended particles size and (if applicable) the dilution of the basic suspension.

Such an operation is aimed at preventing the so-called "network overtraining" [39] and improving the ability to generalize. *Dropout*, due to its random nature, also shows the regulari-zation capabilities. *Dropout* layers play a role partially similar to that of *batchNormalization* in neural networks – both can improve data normalization, generalization capabilities and accelerate the convergence of the learning algorithm. However, this is done at the expense of performing additional calculations, which especially in the case of *dropout* increases the total learning time of the network. We have also observed that in the step of finalizing the network learning algorithm and recalculating statistics there is a significant decrease in the accuracy of network classification. In the works [40,41] some hints can be found suggesting the cause of this phenomenon. In particular, the authors of the paper [40] point to a variance shift, which seems likely in our case. In order to reduce this harmful effect, it is suggested to increase the value of the *MiniBatchSize* parameter [36], which, however, turned unsatisfactory in our case.

The *classif* output layer is preceded by a *softmax* layer. Together they transform the output of the *fc2* layer into a class affiliation vector, where the highest value is interpreted as the affiliation of the input image to the class corresponding to the position in the vector.

The *softmax* [42] layer imposes a *softmax* function on the inputs, ensuring that the vector values are within the 0-1 range and add up to unity. Whereas the *classif* layer [43] calculates the cross entropy loss function, which in the course of training the network acts as a target function to be optimized.

The training – cross entropy loss optimisation – progressed correctly for both single-component and single- plus two-component suspensions sets. The *Accuracy* reached the level of ~95% and the *Loss* of ~0.15. The loss function evolution for the single- plus two-component suspensions set is presented in Fig. 5. Training of the *t7g24* network was a rather

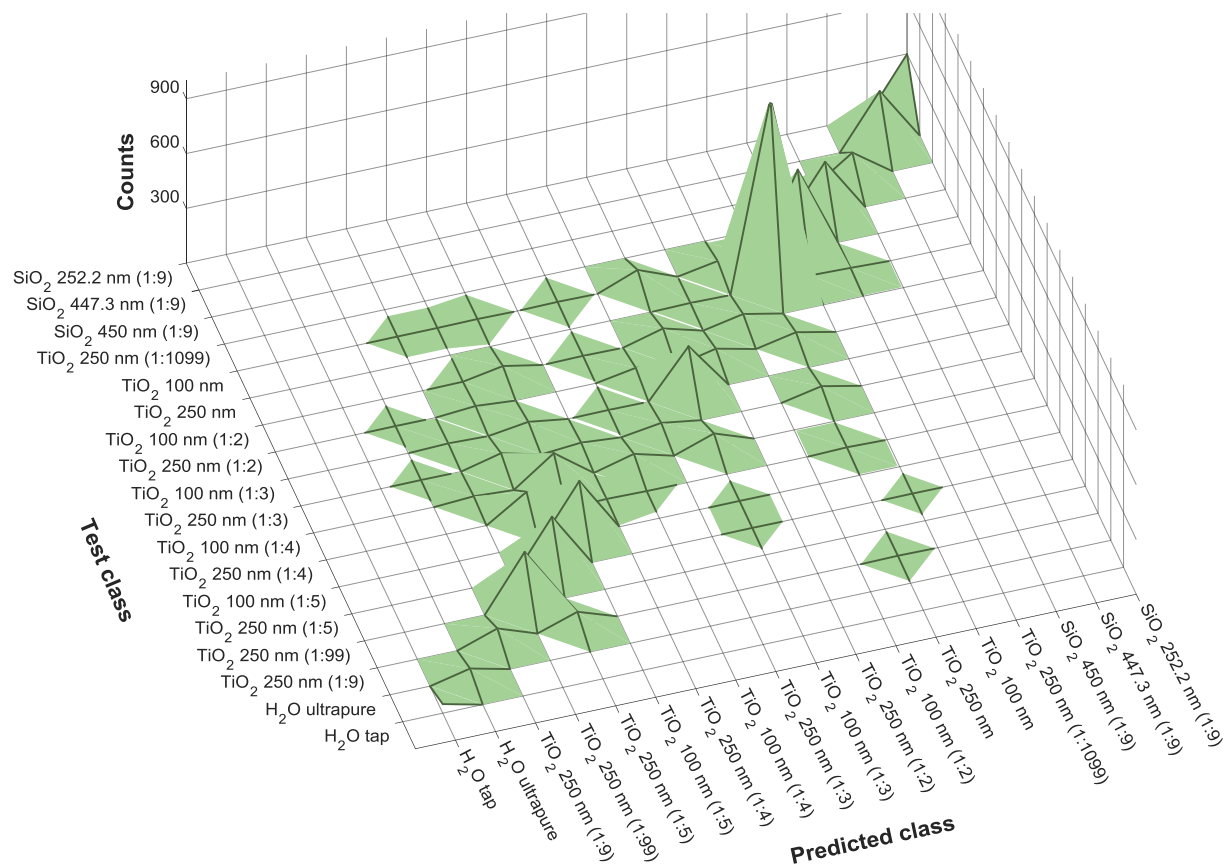

Figure 7. Magnification of the region of maximum confusion of the matrix shown in Fig. 3. All class labels are shown.

time-consuming process of the order of several days (see Fig. 5).

## 4. Results and discussion

In Fig. 6 we present an overview of the classification confusion matrix for a network trained on single-component suspensions. It can be seen that practically all the classes were matched unambiguously, though the similarities between the classes also manifest clearly. It is worth noticing that even the classes which differ only by the dilution were identified properly. However, the dilution seems to be rather difficult to identify, which can be seen in the region of maximal confusion (shown in detail in Fig. 7). Analogously, classes with different particles of the same or similar size often appear similar, as could be expected. The structure of the confusion matrix indicates that the relation of similarity is generally reflexive. There are classes with incomprehensible similarity relations, like *PS 1100 nm* or *Au 100.6 nm*, which might be attributed to some unidentified experimental errors. It must be however underlined that the issue is minor in the case of single-component suspensions.

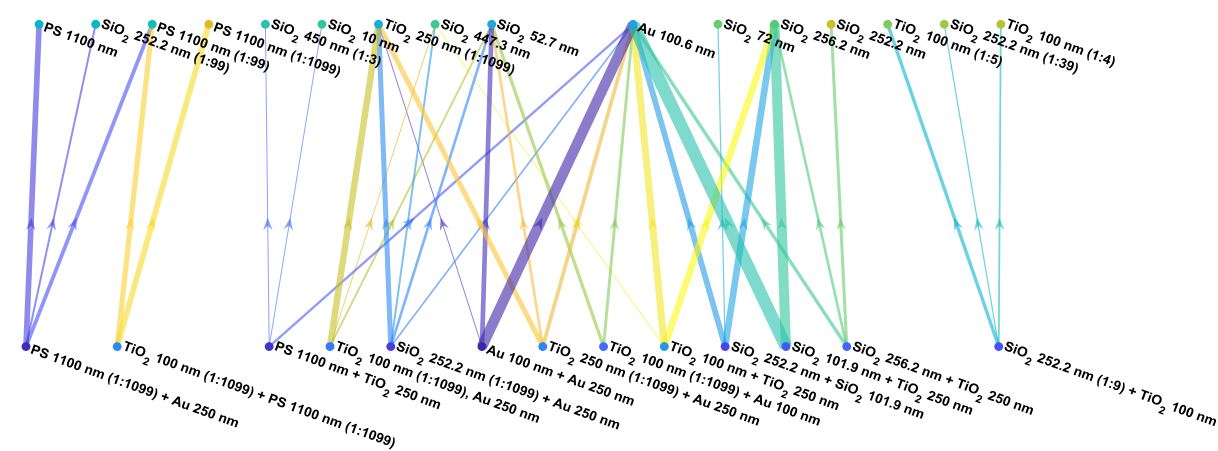

Figure 8. Similarity graph fragment corresponding to the network trained on single-component suspensions (confusion matrix in Figs. 3 and 4) tested with two-component suspensions. Bottom-up representation. The edge thickness corresponds to the number of classification counts.

The network trained on single-component suspensions was then tested on scattering images produced by two-component suspensions. All substances in binary suspensions belonged to the set of single-component suspensions. Such a test should reveal the abstracting ability of the network. The graph showing the matching of two-component suspensions to single-component suspension classes is presented in Fig. 8. The results are promising, but not spectacular. They are obscured by an odd artefact that *Au 100.6 nm* class is similar to very many. However, as it can be read from the graph, two-component mixtures are usually assigned to classes corresponding to at least one of its components. However, it can be observed that the component (class) with larger particles tends to dominate. This is naturally due to the fact that scattering on larger particles is much stronger – for example, from 1100 nm to 100 nm is an order of magnitude of difference in radius, so the scattering is at least 100 times stronger and the scattering image is much brighter. The difference in refractive index between particles of the same size yields a much weaker effect. The information derived from the brighter scattering image – from larger particles – dominates over the information about smaller particles. In order to detect smaller particles in the mixture, an appropriate method would have to be used. A network that detects only one class using a sensitivity threshold might be promising, however it is unclear how it could be integrated with the existing network architecture. As mentioned above, analysing a sequence of images (movie) at once might also help.

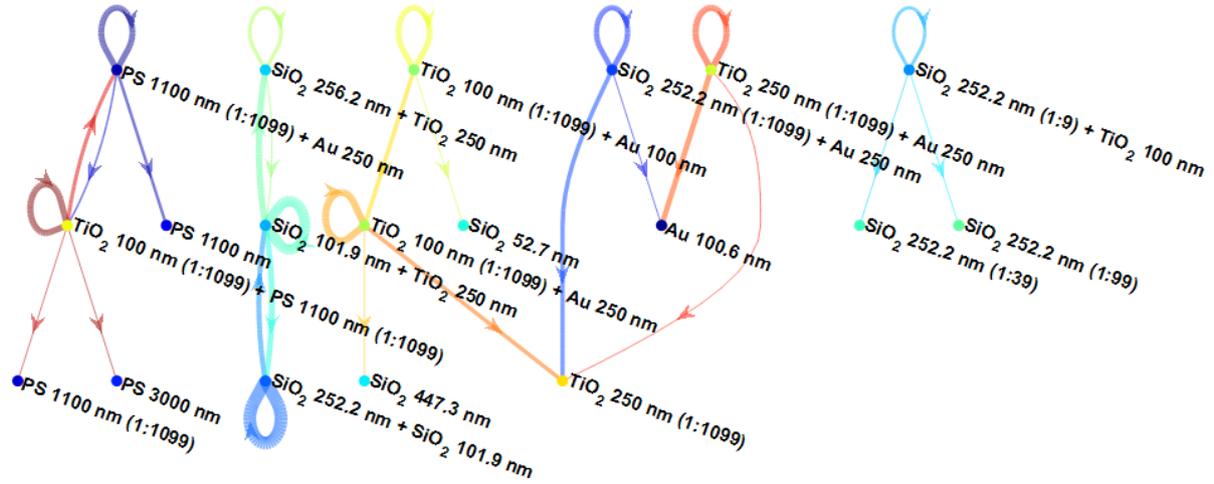

Figure 9. Similarity graph fragment corresponding to the network trained on both single-component and two-component suspensions, tested with two-component suspensions. Top-down representation. The edge thickness corresponds to the number of classification counts.

In the meantime, we performed a different experiment: we trained the network on single- and two-component suspensions at once. The classification confusion matrix for such a network is very similar to that presented in Figs. 6 and 7 and the similarity graph pertaining to two-component suspensions is shown in Fig. 9 . It turned out that the network had learnt to recognize mixtures as such – as separate classes. The two-component classes first of all simply point to themselves and only then point to classes corresponding to their components. A conclusion may be drawn at this point that though the network can easily recognize very many classes of monodisperse suspensions regardless of their dilution, the generalisation capabilities cannot be achieved as a by-product and must be further worked on.

## 5. Conclusions

We showed, to our knowledge – for the first time, that nanoparticle suspensions comprising a multi-parameter set of ~$10^2$ classes, can be quickly identified by classifying the speckle images with a convolution neural network. Since we observed that, for instance, tap water can be accurately distinguished from ultrapure water, it can be inferred that the method is good for any (colloidal) suspension. This is far beyond the capabilities of the human experimenter.

The presented classification system utilised speckle images (movies) from nanoparticle suspensions placed in a thin cuvette. It is expected that classifying suspensions using two-dimensional angular optical scattering (TAOS) images from a microdroplet of suspension might be possible, though will be significantly more difficult due to movements and evolution (size and concentration change) of the whole droplet. Analysing a sequence of images (movie) at once, as speckle dynamics carry information on the suspended particles hydrodynamic radius, can be considered in this context.

The classifier was trained, validated and tested on both single component monodispersive suspensions, as well as on two-component suspensions. It was able to properly recognise all the 73 classes – different suspensions from the training set and shows the capability of learning many more. The classes comprised different nanoparticle material and size, as well as different concentration of suspensions. We also examined the capability of the system to generalise, by testing a system trained on single-component suspensions with two-component suspensions. The capability to generalise was found promising but significantly limited. Pos-sible solutions has been proposed – a network that detects only one class using a sensitivity threshold and/or analysing a sequence of images.

A system using neural network was also compared with the one using support vector machine. SVM was found much more resource-consuming and thus couldn't be tested on full-size images. Using image fragments yields unacceptably poor results for both SVM and neural networks.

The results obtained seem promising for the purpose of automatic recognition of suspensions. The use of a well-trained neural network should enable an initial diagnosis of the suspension in real time.

**Appendix A – Comparison with Support Vector Machine**

The Support Vector Machine (SVM) can, like neural networks, be used for classification. This technique consists in fitting a hyperplane that separates two sets as exactly as possible by means of linear regression. In order to adapt SVM to the classification of images (especially images such as the scattering images described in this work), additional steps are needed.

| Classifier | Accuracy |
|---|---|
| linear SVM | 0.1993 |
| SVM with radial basis function (RBF) kernel | 0.0201 |
| SVM with polynomial kernel | 0.0823 |
| convolution neural network labelled as *t7d16re* in [33] | 0.46 |

Table A.1. Comparison of the accuracy of SVMs with different kernels and a neural network operating on the same image (fragment) size and test batch. "One vs. One" strategy was used for all presented SVMs.

A single SVM separates linearly only two sets, and in our dataset several dozen classes are distinguished. A combination of multiple classifiers is used to classify into multiple classes using binary classifiers. Many strategies are known, but two popular strategies "One vs. One" and "One vs. All" have been tested. Training multiple classifiers instead of just one significantly increases the training time. The basic SVM is linear, but when there is a suspicion of non-linear relationships in the data, a so-called Kernel Trick can be used to add a non-linear transformation. We compared the accuracy of linear SVM with radial basis function (RBF) and polynomial kernels and it turned out that linear SVM works much better than non-linear kernels. This is probably due to the high dimensionality of the problem. The comparison of different SVMs with a comparable neural network is presented in Tab. A.1.

It can be concluded that for (small) image fragments SVM performs only slightly worse than a comparable neural network. However, both perform hardly satisfactory. For a full size image, SVM cannot be implemented, while neural networks perform quite well.

# References

[1] Crosta FG, Pan Y-L, Videen G. Heuristic optical detection of bioaerosols. SPIE Newsroom 2013:10–2. https://doi.org/10.1117/2.1201304.004841.

[2] Aptowicz KB, Pinnick RG, Hill SC, Pan YL, Chang RK. Optical scattering patterns from single urban aerosol particles at Adelphi, Maryland, USA: A classification relating to particle morphologies. J Geophys Res Atmos 2006;111:1–13. https://doi.org/10.1029/2005JD006774.

[3] Holler S, Fuerstenau SD, Skelsey CR. Simultaneous two-color, two-dimensional angular optical scattering patterns from airborne particulates: Scattering results and exploratory analysis. J Quant Spectrosc Radiat Transf 2016;178:167–75. https://doi.org/10.1016/j.jqsrt.2016.01.009.

[4] Onofri FRA, Barbosa S, Touré O, Woźniak M, Grisolia C. Sizing highly-ordered buckyball-shaped aggregates of colloidal nanoparticles by light extinction spectroscopy. J Quant Spectrosc Radiat Transf 2013;126:160–8. https://doi.org/10.1016/j.jqsrt.2012.08.018.

[5] Jakubczyk D, Derkachov G, Kolwas M, Kolwas K. Combining weighting and scatterometry: Application to a levitated droplet of suspension. J Quant Spectrosc Radiat Transf 2013;126:99–104. https://doi.org/10.1016/j.jqsrt.2012.11.010.

[6] Li C, Wu X cheng, Cao J zheng, Chen L hong, Gréhan G, Cen K. Application of rainbow refractometry for measurement of droplets with solid inclusions. Opt Laser Technol 2018;98:354–62. https://doi.org/10.1016/j.optlastec.2017.07.026.

[7] Wilms J, Weigand B. Composition measurements of binary mixture droplets by rainbow refractometry. Appl Opt 2007;46:2109–18.

[8] Goodman JW. Statistical Properties of Laser Speckle Patterns. In: Dainty J.C., editor. Laser Speckle Relat. Phenomena. Top. Appl. Physics, vol. 9, Springer Berlin Heidelberg; 1975, p. 9–75. https://doi.org/10.1007/978-3-662-43205-1_2.

[9] Goodman JW. Statistical optics. NewYork: John Wiley & Sons Inc.; 2000.

[10] Ding C. Convolutional neural networks for particle shape classification using light-scattering patterns. J Quant Spectrosc Radiat Transf 2020;245:106901. https://doi.org/10.1016/j.jqsrt.2020.106901.

[11] Piedra P, Kalume A, Zubko E, Mackowski D, Pan Y Le, Videen G. Particle-shape classification using light scattering: An exercise in deep learning. J Quant Spectrosc Radiat Transf 2019;231:140–56. https://doi.org/10.1016/j.jqsrt.2019.04.013.

[12] Piedra P, Gobert C, Kalume A, Pan Y Le, Kocifaj M, Muinonen K, et al. Where is the machine looking? Locating discriminative light-scattering features by class-activation mapping. J Quant Spectrosc Radiat Transf 2020;247:106936. https://doi.org/10.1016/j.jqsrt.2020.106936.

[13] Kuo CCJ, Zhang M, Li S, Duan J, Chen Y. Interpretable convolutional neural networks via feedforward design. J Vis Commun Image Represent 2019;60:346–59. https://doi.org/10.1016/j.jvcir.2019.03.010.

[14] Wang P, Zhang H, Patel VM. SAR Image Despeckling Using a Convolutional Neural Network. IEEE Signal Process Lett 2017;24:1763–7. https://doi.org/10.1109/LSP.2017.2758203.

[15] Zakharov P, Bhat S, Schurtenberger P, Scheffold F. Multiple-scattering suppression in dynamic light scattering based on a digital camera detection scheme. Appl Opt

2006;45:1756–64. https://doi.org/10.1364/AO.45.001756.

[16] Li Y, Xue Y, Tian L. Deep speckle correlation: a deep learning approach toward scalable imaging through scattering media. Optica 2018;5:1181–90. https://doi.org/10.1364/OPTICA.5.001181.

[17] Redding B, Choma MA, Cao H. Speckle free laser imaging. Conf. Lasers Electro-Optics 2012, vol. i, Washington, D.C.: OSA; 2012, p. JTh2A.86. https://doi.org/10.1364/CLEO_AT.2012.JTh2A.86.

[18] Metzger NK, Spesyvtsev R, Bruce GD, Miller B, Maker GT, Malcolm G, et al. Harnessing speckle for a sub-femtometre resolved broadband wavemeter and laser stabilization. Nat Commun 2017;8:1–8. https://doi.org/10.1038/ncomms15610.

[19] Braga RAJ, Rivera FP, Moreira J. A Practical Guide to Biospeckle Laser Analysis. Theory and Software. Lavras: Ed. UFLA; 2016.

[20] Khodadad D. Combined Digital Holography and Speckle Correlation for Rapid Shape Evaluation (Licentiate dissertation). Luleå University of Technology, 2014.

[21] Kalyzhner Z, Levitas O, Kalichman F, Jacobson R, Zalevsky Z. Photonic human identification based on deep learning of back scattered laser speckle patterns. Opt Express 2019;27:36002. https://doi.org/10.1364/OE.27.036002.

[22] Kürüm U, Wiecha PR, French R, Muskens OL. Deep learning enabled real time speckle recognition and hyperspectral imaging using a multimode fiber array. Opt Express 2019;27:20965. https://doi.org/10.1364/OE.27.020965.

[23] Zhou K, Zhou C, Sapre A, Pavlock JH, Weaver A, Muralidharan R, et al. Dynamic Laser Speckle Imaging Meets Machine Learning to Enable Rapid Antibacterial Susceptibility Testing (DyRAST). ACS Sensors 2020;5:3140–9. https://doi.org/10.1021/acssensors.0c01238.

[24] Gupta RK, Bruce GD, Powis SJ, Dholakia K. Deep Learning Enabled Laser Speckle Wavemeter with a High Dynamic Range. Laser Photon Rev 2020;14:2000120. https://doi.org/10.1002/lpor.202000120.

[25] Woźniak M, Archer J, Wojciechowski T, Derkachov G, Jakubczyk T, Kolwas K, et al. Application of a linear electrodynamic quadrupole trap for production of nanoparticle aggregates from drying microdroplets of colloidal suspension. J Instrum 2019;14:P12007--P12007. https://doi.org/10.1088/1748-0221/14/12/p12007.

[26] Woźniak M, Jakubczyk D, Derkachov G, Archer J. Sizing of single evaporating droplet with Near-Forward Elastic Scattering Spectroscopy. J Quant Spectrosc Radiat Transf 2017;202:335–41. https://doi.org/10.1016/j.jqsrt.2017.08.017.

[27] Jakubczyk D. Thin cuvette speckle sample movies. Mendeley Data 2020. https://doi.org/10.17632/wnmn2pkg2s.1.

[28] Chen H, Kou X, Yang Z, Ni W, Wang J. Shape- and Size-Dependent Refractive Index Sensitivity of Gold Nanoparticles. Langmuir 2008;24:5233–7. https://doi.org/10.1021/la800305j.

[29] McPeak KM, Jayanti S V., Kress SJP, Meyer S, Iotti S, Rossinelli A, et al. Plasmonic Films Can Easily Be Better: Rules and Recipes. ACS Photonics 2015;2:326–33. https://doi.org/10.1021/ph5004237.

[30] Khlebtsov BN, Khanadeev VA, Khlebtsov NG. Determination of the Size, Concentration, and Refractive Index of Silica Nanoparticles from Turbidity Spectra. Langmuir 2008;24:8964–70. https://doi.org/10.1021/la8010053.

[31] Bodurov I, Yovcheva T, Sainov S. Refractive index investigations of nanoparticles dispersed in water. J Phys Conf Ser 2014;558:012062. https://doi.org/10.1088/1742-6596/558/1/012062.

[32] Ma X, Lu JQ, Brock RS, Jacobs KM, Yang P, Hu XH. Determination of complex refractive index of polystyrene microspheres from 370 to 1610 nm. Phys Med Biol 2003;48:4165–72. https://doi.org/10.1088/0031-9155/48/24/013.

[33] Supplementary material for T. Jakubczyk's Masters Thesis n.d. https://github.com/sigrond/MGR.

[34] Pecora R, editor. Dynamic Light Scattering. Applications of Photon Correlation Spectroscopy. New York: Plenum Press; 1985. https://doi.org/10.1007/978-1-4613-2389-1.

[35] Derkachov G, Jakubczyk D, Kolwas K, Piekarski K, Shopa Y, Woźniak M. Dynamic light scattering investigation of single levitated micrometre-sized droplets containing spherical nanoparticles. Meas J Int Meas Confed 2020;158:107681. https://doi.org/10.1016/j.measurement.2020.107681.

[36] Options for training deep learning neural network - MATLAB trainingOptions n.d. https://www.mathworks.com/help/deeplearning/ref/trainingoptions.html.

[37] Hochreiter S. The vanishing gradient problem during learning recurrent neural nets and problem solutions. Int J Uncertainty, Fuzziness Knowledge-Based Syst 1998;6:107–16.

[38] He K, Zhang X, Ren S, Sun J. Deep residual learning for image recognition. Proc. IEEE Conf. Comput. Vis. pattern Recognit., 2016, p. 770–8.

[39] Sjöberg J, Ljung L. Overtraining, regularization, and searching for minimum in neural networks. IFAC Proc Vol 1992;25:73–8.

[40] Li X, Chen S, Hu X, Yang J. Understanding the disharmony between dropout and batch normalization by variance shift. Proc IEEE Comput Soc Conf Comput Vis Pattern Recognit 2019;2019-June:2677–85. https://doi.org/10.1109/CVPR.2019.00279.

[41] Garbin C, Zhu X, Marques O. Dropout vs. batch normalization: an empirical study of their impact to deep learning. Multimed Tools Appl 2020;79:12777–815. https://doi.org/10.1007/s11042-019-08453-9.

[42] Softmax layer - MATLAB n.d. https://www.mathworks.com/help/deeplearning/ref/nnet.cnn.layer.softmaxlayer.html.

[43] Classification output layer - MATLAB classificationLayer n.d. https://www.mathworks.com/help/deeplearning/ref/classificationlayer.html.